\documentstyle[eqsecnum,aps]{revtex}

\input{epsf.tex}
\begin{document}
\draft
\twocolumn[\hsize\textwidth\columnwidth\hsize\csname
@twocolumnfalse\endcsname
\preprint{HEP/123-qed}
\renewcommand{\thefootnote}{\alph{footnote}} %%%%%%%%%%%%%
\title{Chaos and the continuum limit in the gravitational $N$-body problem.
I. Integrable potentials}
\author{Henry E. Kandrup\footnote{Electronic address: kandrup@astro.ufl.edu}}
\address{ Department of Astronomy, Department of Physics, and
Institute for Fundamental Theory
\\
University of Florida, Gainesville, Florida 32611}
\author{Ioannis V. Sideris\footnote{Electronic address: sideris@astro.ufl.edu}}
\address{Department of Astronomy, University of Florida, Gainesville, 
Florida 32611\\}

\date{\today}
\maketitle
\begin{abstract}
This paper summarises a numerical investigation of the statistical properties 
of orbits evolved in `frozen,' time-independent $N$-body realisations of 
smooth, time-independent density distributions corresponding to integrable 
potentials, allowing for $10^{2.5}{\;}{\le}{\;}N{\;}{\le}{\;}10^{5.5}$.
Two principal conclusions were reached: (1) In the limit of
a nearly `unsoftened' two-body kernel, {\it i.e.}, 
$V(r){\;}{\propto}{\;}(r^{2}+{\epsilon}^{2})^{-1/2}$
for small ${\epsilon}$, the value of the largest Lyapunov exponent ${\chi}$ 
does {\it not} appear to decrease systematically with increasing  $N$, so 
that, viewed in terms of the sensitivity of individual orbits to small changes
in initial conditions, there is no sense in which chaos `turns off' for large 
$N$. (2) Nevertheless, there is a clear, quantifiable sense in which, on the 
average, as $N$ increases chaotic orbits in the frozen-$N$ systems come to 
more closely resemble integrable characteristics in the smooth potential. 
When viewed in configuration or velocity space, or as probed by collisionless
invariants like angular momentum, frozen-$N$ orbits typically diverge from 
smooth potential characteristics as a power law in time on a time scale 
${\propto}{\;}N^{p}t_{D}$, with $t_{D}$ a characteristic dynamical, 
or crossing, time. For the case of angular momentum, the divergence is well 
approximated by a $t^{1/2}$ dependence, so
that, when viewed in terms of collisionless invariants, discreteness effects
acts as a diffusion process which, presumably, can be modeled by nearly white
Gaussian noise in the context of a Langevin or Fokker-Planck description. For
position and velocity, the divergence is somewhat more rapid and characterised
by a $t^{q}$ power law growth with $q{\;}{\approx}{\;}1$, a result that likely 
reflects the effects of linear phase mixing.
\end{abstract}
\pacs{PACS number(s): 05.60.+w, 51.10.+y, 05.40.+j}
]
%Valid PACS appear here.
%{\tt$\backslash$\string pacs\{\}} should always be input,
%even if empty.}
\narrowtext
%\twocolumn

\section{INTRODUCTION AND MOTIVATION}
 \label{sec:level1}
Many astronomical objects, including, {\it e.g.,} globular clusters, 
are typically modeled by bulk gravitational potentials which manifest a high 
degree of symmetry and which, being integrable, lead to completely regular 
characteristics with no possibility of chaotic behaviour. One knows, 
however, that such bulk potentials constitute idealisations, the true system 
corresponding (at least approximately) to a realisation of the gravitational 
$N$-body problem. The important point, then, is that motion in the 
$N$-body problem, even for an $N$-body system which samples a smooth,
time-independent phase
space distribution corresponding to an integrable potential, is typically 
chaotic in the sense that orbits exhibit exponential sensitivity towards 
small changes in initial conditions\cite{1}.
This perhaps is not surprising. The true potential associated 
with a collection of point masses no longer possesses the symmetries of the 
original integrable potential, so that there is no reason why the orbits 
should not be chaotic. 

However, what {\it is}, perhaps, surprising is the expectation, derived both 
from theoretical arguments \cite{2,3}
and from numerical simulations\cite{4},
that the $N$-body problem remains chaotic even for
very large $N$. Suppose, {\it e.g.,} that a system of total mass $M=1$ is 
represented by a collection of $N$ objects of mass $m=1/N$, so distributed 
as to sample the density distribution corresponding to an 
integrable potential. The claim then is that, when expressed in units of a 
natural dynamical, or crossing, time $t_{D}{\;}{\sim}\;1/\sqrt{G{\rho}}$,
with ${\rho}$ a typical density, the 
characteristic time scale ${\tau}$ on which an initial perturbation in any 
given orbit tends to grow will not diverge for $N\to\infty$. In this sense,
the degree of chaos manifested by individual orbits is not expected to `turn 
off' for very large $N$. There is an apparent consensus, motivated both from 
theory and numerical experiments, that ${\tau}$ should not increase without 
bound for $N\to\infty$, although there is some disagreement in the literature 
as to whether ${\tau}(N)$ should converge towards an $N$-independent value 
\cite{2} or whether ${\tau}$ should instead slowly {\it decrease} with 
increasing $N$\cite{3}. 

If, however, this be true, one is confronted with subtle questions of 
principle regarding the nature of the continuum limit. It is generally assumed
\cite{5}
that, for sufficiently large $N$, a self-gravitating system of discrete point 
masses can be characterised adequately by a smooth phase space density that 
solves the collisionless Boltzmann equation ({\it CBE}), {\it i.e.,} the
gravitational analogue of the Vlasov equation from plasma physics. The obvious
point, then, is that time-independent solutions to this equation which manifest
a high degree of symmetry correspond typically to bulk potentials which are
integrable or, even if they be nonintegrable, admit large measures 
of regular orbits. But how is one to reconcile integrable or near-integrable
behaviour in such bulk potentials with the presumed fact that, even for very
large $N$, individual orbits in the true $N$-body problem typically manifest 
chaotic behaviour on a time scale ${\sim}{\;}t_{D}$?

Strictly speaking, there is no logical contradiction: It is completely
possible for collective properties of an $N$-body system to be described
correctly by the {\it CBE}, even if the characteristics associated
with the self-consistent potential do not coincide, even approximately, with
real $N$-body trajectories. However, it {\it would} seem important to pin 
down carefully what is actually going on:
\par\noindent ${\bullet}$
Is it really true that individual trajectories in the $N$-body problem are
chaotic for very large $N$, even if the bulk potential associated with the
system is integrable? The indications are that the answer to this is: yes. 
However, most of the work done to date on chaos in the $N$-body problem has 
focused on systems with comparatively small $N$ and/or a hierarchy of masses,
or, for larger systems, on comparatively short time behaviour. Little if any 
work has been
done to provide estimates of  honest Lyapunov exponents over intervals 
${\gg}{\;}t_{D}$ for large $N$ systems comprised of bodies of comparable mass.
\par\noindent ${\bullet}$
Does this chaos reflect simply the fact that the system is grainy, or does
it reflect the details of the full $N$-body dynamics? If, {\it e.g.,} one
were to replace a smooth time-independent, integrable potential by the 
irregular potential associated with an $N$-particle sampling of the smooth 
mass density that is frozen in time, to what extent will motion in that grainy 
time-independent potential be chaotic? 
\par\noindent ${\bullet}$
Even presuming that the $N$-body problem is chaotic on a time scale ${\sim}{\;}
t_{D}$, is there some well defined sense in which the `average' properties of
individual $N$-body trajectories track characteristics given by the {\it CBE}?

These conceptual issues are also related directly to the problem of 
`softening.' It is generally recognised that, for small $N$, close encounters 
between individual masses are more important dynamically than for larger 
$N$\cite{5}. For this reason, $N$-body simulators interested in exploring the 
physics of the $N$-body problem for larger $N$ often suppress the effects of
close encounters artificially by replacing the true $1/r$ potential by a 
softened potential $V(r){\propto}\;(r^{2}+{\epsilon}^{2})^{-1/2}$ for some 
`softening parameter' ${\epsilon}$. This certainly suppresses encounters with
impact parameters $<{\epsilon}$ which, presumably, is a good thing. However,
there are strong indications\cite{6} that orbits integrated with such a 
softened potential tend to be `less chaotic' in their behaviour, so that the 
introduction of softening also has the potentially undesirable effect of 
removing $N$-body chaos which really ought to be present, even for very large 
$N$. In any event, earlier investigations of chaos in the $N$-body problem 
based on simulations which incorporate a large amount of softening must be 
viewed with suspicion, since such simulations could 
suppress precisely the effects which one might wish to explore!

This paper summarises a detailed exploration of chaos in time-independent 
potentials generated by sampling the smooth density ${\rho}({\bf r})$
associated with a time-independent solution to the {\it CBE} to create a 
frozen $N$-body realisation of that equilibrium. Most of the work focuses on 
the particularly simple case of an integrable Plummer potential\cite{5}, which 
derives from a spherically symmetric mass distribution. However, it was also
confirmed that, modulo one point discussed in the concluding section, the same 
qualitative results obtain for the potential
associated with a constant density spherical configuration.

Section II begins by describing the numerical experiments that were performed.
Section III summarises a computation of {\it honest} Lyapunov exponents in 
frozen $N$-body realisations of the Plummer potential, exploring how the 
largest exponent ${\chi}$ associated with representative initial conditions 
varies as a function of ${\epsilon}$ and $N$. The principal conclusion here is 
that, at least for small values of ${\epsilon}$, orbits in such potentials are
invariably chaotic; and that, even for particle number as large as 
$N=10^{5.5}$, there is no sense in which increasing $N$ `turns the chaos off.'
Section IV demonstrates that, even though the Lyapunov exponents do not 
decrease with increasing $N$, there is a well defined sense in which, as $N$
increases, orbits in frozen-$N$ potentials remain `close' to smooth potential
characteristics with the same initial condition for progressively longer times.
Section V concludes by summarising the principal conclusions, providing a 
simple physical interpretation, and then commenting on potential implications.

The principal conclusion of this paper is that, for integrable smooth 
potentials which admit no chaos, the continuum limit makes sense even at the 
level of pointwise properties of individual trajectories. The possibility of
chaotic characteristics leads necessarily to very different behaviour and,
for this reason, the case of nonintegrable potentials that admit both regular
and chaotic characteristics will be considered in a separate paper.

\section{DESCRIPTION OF THE NUMERICAL EXPERIMENTS}

\par\noindent
The numerical computations reported here were performed for a so-called 
Plummer potential,
\begin{equation}
{\Phi}(r)=-{GM\over \sqrt{r^{2}+b^{2}}}.
\end{equation}
This potential is generated via Poisson's equation from a density profile
\begin{equation}
{\rho}(r)=\left({3M\over 4{\pi}b^{3}}\right)
\left(1 + {r^{2}\over b^{2}}\right)^{-5/2},
\end{equation}
and corresponds to an equilibrium solution to the {\it CBE} satisfying
\begin{equation}
f(E)=\cases{ A(-E)^{7/2} & if ${\Phi}(r=0)<E={1\over 2}v^{2}+{\Phi}<0$; \cr
\;\;\;\;\;\;	0        & if $\;\;\;\;\;\;\;\;\;\;\;\;\;\;\;\;\;\;E={1\over 2}v^{2}+{\Phi}>0$.\cr}
\end{equation}
Units were so chosen that $G=M=b=1$.

The principal aim was to compare orbits generated in the smooth potential
with orbits evolved in time-independent $N$-body realisations of the 
potential. For a variety of fixed values of $N$ and ${\epsilon}$, $20$ 
different time-independent $N$-body potentials were constructed. Each of these 
was associated with a random sampling of the smooth density distribution
generated using a von Neumann rejection algorithm (cf. \cite{7}). 
This entailed constructing singular density distributions 
\begin{equation}
{\rho}_{N}({\bf r})={M\over N}\,\sum_{i=1}^{N}{\delta}_{D}({\bf r}-{\bf r}_{i})
,
\end{equation}
which, allowing for a softening parameter ${\epsilon}$, yielded potentials
of the form 
\begin{equation}
V_{N}({\bf r})=-{GM\over N}\,\sum_{i=1}^{N}\,{1
\over \sqrt{({\bf r}-{\bf r}_{i})^{2}+{\epsilon}^{2}}}.
\end{equation}
The objective then was to select individual initial conditions $({\bf r}_{0},
{\bf v}_{0})$ and to evolve these initial conditions in both the smooth
potential and the $20$ `frozen' $N$-body potentials, while simultaneously
tracking the evolution of a small initial perturbation, periodically 
renormalised at fixed intervals ${\delta}t$, so as to extract an
estimate of the largest (short time) Lyapunov exponent \cite{8}.

The integrations were performed for a time corresponding physically to 
${\sim}{\;}100t_{D}$ using a Runge-Kutta integrator that typically conserved
energy to at least one part in $10^{4}$. The value $100t_{D}$ was selected (i) 
because it corresponded to an interval sufficiently long that one began to see
convergence towards a well-defined Lyapunov exponent ${\chi}$ and, perhaps
more importantly, (ii) because, for physical systems like real galaxies,
$100t_{D}$ corresponds to an interval comparable to the age of the Universe.

The total particle number $N$ in the `frozen' $N$-body potentials was allowed 
to vary between $N=10^{2.5}$ and $N=10^{5.5}$. Physical interest focuses
primarily on the limit ${\epsilon}\to 0$, this corresponding to an `honest'
$N$-body calculation. However, the effects of a nonzero ${\epsilon}$ were
also considered in some detail, with the aim of ascertaining possible 
undesirable consequences for conventional $N$-body simulations, which 
typically involve a substantial softening. The experiments with variable 
${\epsilon}$ indicated that, for ${\epsilon}<10^{-4}$ or so, the precise value 
of ${\epsilon}$ was largely immaterial, at least statistically. 

\section{SHORT TIME LYAPUNOV EXPONENTS}
The principal diagnostic here was the mean (short time) Lyapunov exponent
${\langle}{\chi}{\rangle}$, generated, for a given choice of initial condition
and for specified values of ${\epsilon}$ and $N$, as the average value of
${\chi}$ at $t=100t_{D}$ for $20$ different frozen-$N$ potentials. The 
fundamental question was how, for fixed initial
condition, this ${\langle}{\chi}{\rangle}$ depends on ${\epsilon}$ and $N$.
FIGURE 1 exhibits ${\langle}{\chi}{\rangle}$ as a function of $\log_{10}
{\epsilon}$ for multiple integrations of one representative initial condition, 
which corresponded initially
to a roughly isotropic distribution of velocities, allowing for several 
different values of $N$. FIGURE 2 gives ${\langle}{\chi}{\rangle}$ as a 
function of $\log_{10} N$ for the same initial condition, now allowing for 
several different values of ${\epsilon}$.

It is evident from FIGURE 1 that, at least for comparatively large values of
softening parameter,
decreasing ${\epsilon}$ tends to make the orbit more chaotic. This is hardly
surprising: Since the smooth potential is integrable, one anticipates that
the chaos is associated completely with close encounters between the test
mass and individual frozen masses. The introduction of a nonzero smoothing
corresponds {\it de facto} to the introduction of a minimum impact parameter
(since the potential is bounded in magnitude by $V_{max}=-GM/N{\epsilon}$)
but the existence of such a minimum impact parameter limits the maximum effect
that can arise from a close encounter. 

However, for sufficiently small values of ${\epsilon}$, the precise value
of ${\epsilon}$ appears to be largely immaterial. This again is hardly 
surprising: As long as ${\epsilon}$ is small compared with the value of the
closest separation between the test particle and any of the frozen particles
during the course of the integration, the test particle feels an essentially
unsoftened potential and should behave (at least statistically) as if
${\epsilon}=0$. The point then is that, for $N{\;}{\le}{\;}10^{6}$ and an 
integration
time as short as $100t_{D}$, the minimum separation associated with the 
closest encounter between the test mass and 
any of the frozen masses should be greater than or comparable to 
${\epsilon}{\;}{\sim}{\;}10^{-4}$.
Indeed, a simple geometric argument indicates\cite{9} that the time scale
$t_{\epsilon}$ for a close encounter with minimum separation as small as
${\epsilon}$ scales as
\begin{equation}
{t_{\epsilon}\over t_{D}}{\;}{\sim}{\;}{R_{sys}^{2}\over N{\epsilon}^{2}},
\end{equation}
where $R_{sys}$ is the size of the system in question.

One obvious implication of these results is that the introduction of a large
amount of softening into a numerical simulation can have the unnatural result 
of significantly decreasing the amount of chaos manifested by individual orbits
in a real astronomical system.

For comparatively large values of ${\epsilon}$, ${\langle}{\chi}{\rangle}$ 
decreases rapidly with increasing $N$ but, for sufficiently small values of 
$N$, it appears that ${\langle}{\chi}{\rangle}$ is nearly independent of $N$
(although there are hints that ${\langle}{\chi}{\rangle}$ may continue to 
{\it increase} very slowly). The fact that, for large ${\epsilon}$, 
${\langle}{\chi}{\rangle}$ 
should decrease with increasing $N$ can again be explained by comparing the
magnitude of ${\epsilon}$ with the typical distance between masses in the 
system, which is of order $n^{-1/3}{\;}{\sim}{\;}R_{sys}/N^{1/3}$, with $n$ 
a characteristic number density. If ${\epsilon}$ is larger than, or comparable
to, $n^{-1/3}$, even weak close encounters are essentially `turned off,' so 
that the source of chaos has been largely reduced, if not completely removed.
The fact that ${\langle}{\chi}{\rangle}$ should be essentially independent of 
$N$ in the limit ${\epsilon}\to 0$ has been argued by various authors in a
number of different ways\cite{2,3}. A simple physical explanation is provided
in the concluding section.

If a single orbit be integrated for progressively longer times, how quickly
will the short time Lyapunov exponent ${\chi}(t)$ converge towards the true
time-independent ${\chi}$? Studies of orbits in smooth nonintegrable potentials
reveal that, when the phase space is highly complex and, because of the
Arnold web, orbits can be `stuck' temporarily in regions where the short time
Lyapunov exponents are especially small or especially large, the time required
for a reasonable level of convergence can be extremely long,
${\sim}{\;}10^{5}t_{D}-10^{6}t_{D}$ or even more\cite{8}. If, however, the
phase space is simpler in the sense that the Arnold web forms less of an
impediment and such trapping is comparatively rare, the time required is 
typically much shorter. One way in which to quantify the overall rate of 
convergence is by performing a simple time series analysis: An orbit segment 
of length $T$ can of course be divided into $k$ segments of length 
${\Delta}t=T/k$ and a short time Lyapunov exponent ${\chi}({\Delta}t)$ 
computed for each segment. The dispersion
${\sigma}_{\chi}({\Delta}t)$ then provides a useful probe of the degree to
which, on time scales ${\sim}{\;}{\Delta}t$, the degree of chaos exhibited
by different orbit segments is more or less the same. Determining
${\sigma}_{\chi}$ as a function of ${\Delta}t$ provides a quantitative
characterisation of the rate of convergence towards a unique ${\chi}_{\infty}$.
A simple argument based on the Central Limits Theorem suggests\cite{10} that, 
if the 
accessible phase space regions are simple and trapping is rare, so that the
amounts of chaos manifested at times $t$ and $t+{\Delta}t$ are essentially
uncorrelated, 
\begin{equation}
{\sigma}_{\chi}{\;}{\propto}{\;}({\Delta}t)^{-p},
\end{equation}
with $p=1/2$. If, alternatively, the phase space is complex
and trapping is important, one would expect that ${\sigma}_{\chi}$ decreases
much more slowly with increasing ${\Delta}t$.

Such a time series analysis was performed for the data sets used to generate
the mean exponents ${\langle}{\chi}{\rangle}$. For each set of $20$ 
integrations, each orbit segment of length $T=100t_{D}$ was separated into
$k$ segments of length ${\Delta}t=T/k$. A short time Lyapunov exponent
${\chi}({\Delta}t)$ was then computed for each of the resulting $20k$ segments,
and these were used to compute the dispersion ${\sigma}_{\chi}({\Delta}t)$.
Allowing for $k=2^{q}$, for $q=0,1,2,3,4,5,$ and $6$ was equivalent to
varying ${\Delta}t$ between ${\Delta}t=(100/64)t_{D}$ and ${\Delta}t=100t_{D}$.
This time series analysis led to the conclusion that the dispersion
${\sigma}_{\chi}$ is typically well fit by a power law dependence of the form
given by eq. (3.2), although the exponent $p$ tends to be somewhat smaller 
than $p=1/2$, the best fit value typically satisfying $p{\;}{\sim}{\;}0.4$.
Several examples are exhibited in FIGURE 3.
The fact that $p$ is comparatively close to $1/2$, rather than the much
smaller values that are often observed in very `sticky' nonintegrable 
potentials\cite{10}, corroborates the intuition that, because the chaos in
this problem is associated exclusively with close encounters, trapping is
rare and the degree of chaos exhibited at different times tends to be 
statistically uncorrelated.

\section{COMPARISON OF SMOOTH AND N-BODY ORBITS}

It is clear that, for sufficiently short times, a frozen-$N$ orbit will
coincide almost exactly with the smooth potential characteristic associated
with the same initial condition. And similarly, it is clear that, at 
sufficiently late times, the irregularities in the frozen-$N$ potential 
will cause the frozen-$N$ orbit to deviate significantly from the smooth
characteristic. Probing the validity of the continuum limit at the level of
individual orbits thus devolves into determining the rate at which the 
frozen-$N$ orbits and smooth characteristics diverge. In this connection, two
obvious questions arise. Do frozen-$N$ orbits diverge from the smooth 
characteristics exponentially or as a power law in time? And how does the
overall rate of divergence depend on $N$?

Such probes of the validity of the continuum limit differ from the ordinary
point of view, where convergence is typically defined in terms of quantities
like bulk moments of the system, ignoring completely the behaviour of 
individual trajectories. A possible intermediate characterisation is to
focus {\it not} on the pointwise behaviour of the chaotic orbits but, instead,
on quantities which might be less sensitive to the $N$-body chaos. In 
particular, one can also ask: How do frozen-$N$ orbits deviate from smooth 
characteristics in terms of quantities which, in the smooth potential, 
correspond to time-independent constants of the motion, like angular momentum 
in a spherically symmetric system?

These questions were addressed here both visually and quantitatively through
a computation of the statistical properties of frozen-$N$ orbits. Given $n=20$
different trajectories $\{({\bf r}_{i}(t),{\bf v}_{i}(t))\}$, $i=1,...,n$,
and the smooth characteristic $({\bf r}_{s}(t),{\bf v}_{s}(t))$ associated 
with the same initial condition, there are two types of moments which one 
might choose to consider. Quantities like 
\begin{equation}
{\langle}{\bf r}{\rangle}={1\over n}\sum_{i=1}^{m}\,{\bf r}_{i}
\end{equation}
and
\begin{equation}
Dr^{2}={\langle}|{\bf r}_{i}-{\langle}{\bf r}{\rangle}|^{2}{\rangle}
{\;}{\equiv}{\;}
{1\over n}\sum_{i}\,|{\bf r}_{i}-{\langle}{\bf r}{\rangle}|^{2}
\end{equation}
and the corresponding quantities generated from ${\bf v}$ and 
${\bf J}={\bf r}{\times}{\bf v}$ focus on the frozen-$N$ orbits in and of 
themselves. Alternatively, such moments as
\begin{equation}
{\Delta}r^{2}{\;}{\equiv}{\;}{\langle}|{\bf r}-{\bf r}_{s}|^{2}{\rangle}=
{1\over n}\sum_{i}\,|{\bf r}_{i}-{\bf r}_{s}|^{2}
\end{equation}
and
\begin{equation}
{\delta}r^{2}={\langle}|{\langle}{\bf r}{\rangle}-{\bf r}_{s}|^{2}{\rangle}
\end{equation}
compare the frozen-$N$ orbits with the smooth potential characteristic and,
as such, their behaviour as a function of $N$ is particularly relevant in 
understanding the continuum limit. 
Overall, the quantities $Dr$, ${\delta}r$, and ${\Delta}r$ were found to 
exhibit comparatively similar evolutions, so that attention below focuses 
on the moments ${\langle}{\bf r}{\rangle}$ and ${\Delta}r$, which 
seem especially natural physically.

The most striking conclusion is that individual frozen-$N$ orbits typically
diverge from the smooth characteristic as a power law in time, {\it not}
exponentially. This is true both for comparatively large values of 
${\epsilon}$, where the frozen-$N$ orbits are nearly regular, and for smaller
values of ${\epsilon}$, where the orbits are much more chaotic. This result
is perhaps surprising. One might naively have supposed that, since the 
frozen-$N$ orbits are strongly chaotic, at least for small ${\epsilon}$, they 
would tend to diverge exponentially from the smooth characteristics on a time
scale ${\tau}{\;}{\sim}{\;}{\chi}^{-1}$. However, such an exponential 
divergence is most definitely {\it not} observed. 

For how long does this power law divergence persist? Does it cease when the 
distance between the frozen-$N$ orbit and the smooth characteristic is still 
small, or does the divergence continue until the frozen-$N$ orbit and the 
smooth characteristic tend to be widely separated in configuration space? 
If, {\it e.g.,} this divergence terminated at comparatively small separations,
much smaller than the size of the system, one could argue that, even though
the frozen-$N$ orbits are chaotic, they still remain `close' to the smooth
characteristics. The answer
here is that this divergence continues until the typical separation has become 
comparable to the size of the configuration space region to which the orbits
are confined and the frozen-$N$ orbit has become completely `decorrelated'
in appearance from the smooth potential characteristic. 

The same conclusion also obtains if one focuses on an ensemble of $20$ 
frozen-$N$ orbits and probes their statistical properties. The six panels of 
FIGURE 4, generated for the initial condition used in FIGURES 1 - 3,
compare ${\langle}x{\rangle}$ for frozen-$N$ ensembles with the smooth
$x_{s}$ for orbits evolved with ${\epsilon}=10^{-5}$, allowing for six values
of $N$ extending from $N=316$ to $N=100 000$. In each case, one finds that,
for sufficiently large $t$, ${\langle}x{\rangle} \to 0$, as would be expected
if the frozen-$N$ orbits have become completely different from one another 
and move through configuration space with random orientations. FIGURE 5 
compares the radial coordinates ${\langle}r{\rangle}$ and $r_{s}$ for the
extreme case of an initial condition corresponding in the smooth potential to
a purely radial orbit. 

The time scale $t_{G}$ on which the frozen-$N$ orbits diverge from the smooth
characteristic, and hence the time scale on which ${\Delta}r$ grows,
increases with increasing $N$. Even though the frozen-$N$ orbits remain
`equally chaotic' in the sense that their Lyapunov exponents ${\chi}$ remain
nearly constant, they remain close to the smooth characteristic for 
progressively longer times. 

The left hand panels of FIGURE 6 exhibit 
${\Delta}r/2^{1/2}R_{s}$, 
with $R_{s}^{2}$ the mean value of $r^{2}$ associated with the 
smooth characteristic, as computed for the same initial condition evolved with 
${\epsilon}=10^{-4}$ for $N=1000$ and $N=100000$. The right hand side exhibits
the same data, recorded at intervals of $0.025t_{D}$, once they have been
subjected to a boxcar averaging over an interval ${\delta}t=1t_{D}$. The
large envelops associated with the curves in the left hand panels reflect,
{\it e.g.,} the fact that, at late times, individual orbits in the $n=20$
orbit ensembles are oscillating about a value of unity. 

That ${\Delta}r/2^{1/2}R_{s}$ converges towards unity is a reflection
of the fact that the orbits have indeed become completely different one from
another: Given that the frozen-$N$ orbits conserve energy, one might expect
that their average distance from the origin should, on the average, be the
same as for the smooth characteristic, so that
\begin{equation}
{\langle}r^{2}(t){\rangle} \to R_{s}^{2} \qquad {\rm for} 
\qquad t\to\infty.
\end{equation} 
Assuming, however, that this be true and that 
\begin{equation}
{\langle}{\bf r}(t){\cdot}{\bf r}_{s}(t){\rangle} \to 0 \qquad {\rm for}
\qquad t\to\infty,
\end{equation}
one infers that ${\Delta}{\bf r}^{2}\to 2\,R_{s}^{2}.$ 
Analogous behaviour is observed for the quantity 
${\Delta}v/2^{1/2}V_{s}$, 
with $V_{s}^{2}$ defined correspondingly for the smooth characteristic.

As is manifested by FIGURE 6, the growth of ${\Delta}r$ and ${\Delta}v$ is
roughly linear in time. Indeed, when comparing an ensemble of frozen-$N$ 
orbits with a smooth orbit characteristic generated from the same initial 
condition, one finds that, for small ${\epsilon}$, ${\Delta}r$ and ${\Delta}v$ 
are both reasonably well fit by a linear growth law of the form
\begin{equation}
{{\Delta}r\over r_{s}}=
{{\Delta}v\over v_{s}}={t\over t_{G}}.
\end{equation}
The growth time $t_{G}$, which is the same for both ${\Delta}r$ and 
${\Delta}v$, satisfies
\begin{equation}
t_{G}{\;}{\approx}{\;}A_{G}N^{p}\,t_{D},
\end{equation}
with $A_{G}$ of order unity and $p{\;}{\approx}{\;}1/2$. 
FIGURE 7 exhibits $\log_{10} (t_{G}/t_{D})$ as a function of $\log_{10} N$ for 
two different initial conditions evolved with ${\epsilon}=10^{-5}$.

The fact that $t_{G}$ scales as $N^{1/2}$ would suggest that the divergence
of the frozen-$N$ orbits from smooth characteristics reflects a diffusion
process, associated with a collection of random close encounters. However,
this might in turn suggest that ${\Delta}r$ and ${\Delta}v$ should grow as
$t^{1/2}$, rather than the approximately linear growth that was observed in
the numerical simulations. Interesting, though, such a $t^{1/2}$ behaviour 
{\it does} obtain for quantities like angular momentum, which are conserved 
absolutely in the smooth potential. Indeed, one finds that, for small 
${\epsilon}$, ${\Delta}J$ satisfies
\begin{equation}
{{\Delta}J^{2}\over J_{s}^{2}}={t\over t_{J}},
\end{equation}
where $J_{s}$ is the typical magnitude of the angular momentum associated
with a characteristic with the specified energy. The growth time $t_{J}$ 
again scales as $N^{1/2}$, but tends to be somewhat larger than $t_{G}$, so
that
\begin{equation}
t_{J}{\;}{\approx}{\;}A_{J}N^{p}\,t_{D},
\end{equation}
with $A_{J}{\;}{\sim}{\;}3A_{G}$ and $p{\;}{\approx}{\;}1/2$. 
FIGURE 8 exhibits $\log_{10} (t_{J}/t_{D})$ as a function of $\log_{10} N$
for the same integrations used to generate FIGURE 7.

There is also a clear visual sense in which, as $N$ increases, the frozen-$N$ 
orbits become progressively more regular in appearance. This is, {\it e.g.,}
evident in FIGURE 9, which exhibits the $x$-$y$ projections of representative
frozen-$N$ orbits with $N$ varying between $N=316$ and $N=316228$, all 
generated from the same initial condition and integrated for a time $t=25t_{D}$
with ${\epsilon}=10^{-5}$. The final panel exhibits the smooth characteristic
associated with the same initial condition. The most obvious point is that, as 
$N$ increases, the configuration space region to which the orbit is restricted 
more closely coincides with the region occupied by the characteristic. For
example, only for the three largest values of $N$ is the orbit `centrophobic'
in the same sense as the characteristic. 

Also evident is the fact that the orbit `looks smoother' for larger values of 
$N$. This visual impression
reflects the fact that, as $N$ increases, the power associated with the 
Fourier spectrum of an orbit tends to become more concentrated near a few 
special frequencies. (Since the smooth orbit associated with the same initial 
condition is regular, all its power is concentrated at a countable set of 
discrete frequencies.) This trend is illustrated in the eight panels of FIGURE 
10,
each of which exhibits $|x({\omega})|$ for a single frozen-$N$ orbit generated 
from the same initial condition. In each case, the data are so normalised that 
the peak frequency has $|x({\omega})|=1$. The spectra were generated from a
time series of $4001$ points, recorded at intervals of $0.025t_{D}$.

The degree to which the orbits become more nearly regular with increasing $N$
can be quantified by determining\cite{11} the `complexity' of the
orbits, {\it i.e.}, the number of frequencies in the discrete Fourier spectrum
which contain an appreciable amount of power. Two such measures of complexity 
are illustrated in FIGURE 11, which was computed for ensembles of frozen-$N$
orbits with varying $N$, all evolved with ${\epsilon}=10^{-5}$ and generated 
from the same
initial condition. The solid curve exhibits $f_{0.1}$, defined as the sum
of the numbers of frequencies $f_{0.1,x}$, $f_{0.1,y}$, and $f_{0.1,z}$
which have more than 10\% as much power as the peak frequencies for
${\omega}_{x}$, ${\omega}_{y}$, and ${\omega}_{z}$, {\it i.e.,}
\begin{equation}
f_{0.1}=f_{0.1,x}+f_{0.1,y}+f_{0.1,z}.
\end{equation}
The dashed curve curve
exhibits $k_{0.95}$, defined correspondingly as the sum of the numbers of 
frequencies required to capture 95\% of the power in the $x$-, $y$-, and 
$z$-directions. In each case, the curve represents an average over different
orbits in the ensemble, and the error bars represent the associated 
dispersions. The obvious point is that both these quantities decrease with
increasing $N$. 

The fact that, as $N$ increases, power becomes more concentrated near a few
special frequencies has important implications for various physical processes
which rely on resonances. For example, a variety of recent arguments in both
galactic and solar system dynamics invoke a process of so-called of 
`resonant relaxation,'\cite{12} which relies on the assumption that, in the
presence of a large central object (a supermassive black hole in the center 
of a galaxy or the Sun at the center of the solar system), $N$-body orbits 
behave very nearly as if they were Keplerian trajectories in the fixed $1/r$ 
potential associated with the central object. If the chaos exhibited by 
individual orbits\cite{13} implied that these orbits were highly irregular,
so that their power was not concentrated near the special Keplerian 
frequencies, resonant relaxation might seem quite implausible. Given, however,
that the orbits become progressively more regular for increasing $N$, resonant
relaxation would seem eminently reasonable, at least for systems in which $N$
is sufficiently large.

\section{CONCLUSIONS AND DISCUSSION}

Even though trajectories remain chaotic in the sense that the largest Lyapunov
exponent does not decrease towards zero, there is a clear sense in which, for 
increasing $N$, orbits in frozen-$N$ potentials exhibit a pointwise convergence
towards characteristics in the smooth potential which the frozen-$N$ 
potentials were chosen to sample. Viewed in configuration or velocity space,
frozen-$N$ orbits tend to diverge linearly from the smooth characteristic with 
the same initial condition on a time scale $t_{G}$ that is proportional to 
$N^{1/2}$. In this sense, the continuum limit appears justified even at 
the level of individual trajectories. The fact that frozen-$N$ orbits remain 
chaotic for very large $N$ is completely consistent with the existence of a 
well-defined continuum limit.

It is easy to understand qualitatively why the frozen-$N$ orbits should remain 
chaotic even for very large $N$. Given that the chaos disappears completely in 
the continuum limit, where the orbits reduce to integrable characteristics, it 
would seem clear that the chaos must be associated with a sequence of `random' 
interactions between a `test' particle and a collection of `field' particles. 
However, this would suggest that the time scale associated with the growth of 
a small initial perturbation can be estimated by considering the tidal effects 
associated with a pair of particles separated by a distance comparable to the 
typical interparticle separation. This tidal acceleration will of course scale 
as
\begin{equation}
{\delta}{\ddot{\bf r}}=({\delta}{\bf r}{\cdot}{\nabla}){\bf a}{\;}{\sim}{\;}
{Gm\over r^{3}}{\delta}{\bf r},
\end{equation}
with $r$ the separation and $m$ the particle mass. Given, however, that
$r{\;}{\sim}{\;}n^{-1/3}{\;}{\sim}{\;}N^{-1/3}R_{sys}$, with $n$ a 
characteristic number density and $R_{sys}$ the size of the system, it follows 
that the time scale ${t_{*}}$ associated with the interaction should satisfy
\begin{equation}
t_{*}{\;}{\sim}{\;}1/\sqrt{G{\rho}}.
\end{equation}
In other words, the time scale associated with any orbital instability induced 
by the graininess of the system should be comparable to the dynamical time 
$t_{D}$, independent of particle number $N$. As $N$ increases, the size of the 
individual particle mass, $m$, and the cube of the typical separation between 
particles, ${\sim}{\;}n^{-1}$, both decrease as $N^{-1}$ so that their ratio 
is independent of particle number.

The fact that the chaotic frozen-$N$ orbits appear to become ``more nearly
regular'' as $N$ increases is consistent with recent claims that the ``scale''
associated with $N$-body chaos decreases with increasing $N$. Specifically,
by comparing trajectories associated with two nearby initial conditions 
evolved in the same frozen-$N$ potential, Valluri and Merritt\cite{14} found
that, when scaled in terms of $R_{sys}$, the size of the system, the typical
separation $R_{sat}$ on which the initial exponential divergence saturates
decreases with increasing particle number, so that $R_{sat}/R_{sys}$ is a 
decreasing function of $N$. 

That the rate of divergence in the nonlinear regime slows more and more for
larger $N$ can be quantified by tracking the actual evolution of two orbits
generated from nearby initial conditions and determining the time required
before their separation becomes `macroscopic.' The result of such an 
investigation is illustrated in FIGURE 12, which was generated once again
from ensembles of $20$ frozen-$N$ orbits all evolved with ${\epsilon}=10^{-5}$.
In each case, the unperturbed orbits were identical to those used to generate
FIGURE 1; the perturbed orbits involved changing the initial value of $x$ by
an amount ${\delta}x=10^{-6}$. FIGURE 12 exhibits as a function of $N$ the
mean time ${\tau}$ required before the separation
\begin{equation}
{\delta}r=({\delta}x^{2}+{\delta}y^{2}+{\delta}z^{2})^{1/2}
\end{equation}
had achieved the value ${\delta}r=1$. (For this initial condition, the
average value of $r$ associated with the smooth characteristic was
$R_{s}{\;}{\approx}{\;}1.83$.) The error bars were derived by
considering the first and second ten orbits in the ensemble separately. 
Because individual orbits diverge at vastly different rates, the dispersion 
associated with a 20 orbit ensemble is much larger than reflected by these 
error bars. It is clear that ${\tau}$ increases systematically 
with increasing $N$, although considerably more slowly than with the $N^{1/2}$
dependence observed for the divergence time scales $t_{G}$ and $t_{J}$.

When viewed in terms of collisionless invariants like angular momentum, 
the divergence of frozen-$N$ orbits from smooth characteristics with the
same initial condition is well approximated as a diffusion process, in which
${\Delta}J$ grows as $(t/t_{J})^{1/2}$ and where, for fixed $t_{D}$, the
divergence time scale $t_{J}$ varies at least approximately as $N^{1/2}$. This 
reinforces the conventional wisdom\cite{15} that discreteness effects may be 
modeled as white, or nearly white, Gaussian noise in the context of a Langevin 
or Fokker-Planck description. It might, therefore, seem somewhat
surprising that, although the divergence time scale $t_{G}$ in configuration 
or velocity space again scales as $N^{1/2}$, the quantities ${\Delta}r$ and 
${\Delta}v$ grow {\it linearly} in time, rather than as $t^{1/2}$.

In this regard, it is significant that if the smooth Plummer potential be
replaced by the smooth potential associated with a constant density
configuration, the linear growth exhibited ${\Delta}r$ and ${\Delta}v$ is in 
fact replaced by the `expected' diffusive behaviour. In this case ${\Delta}r$
and ${\Delta}v$ both grow as $t^{1/2}$, and, when expressed in units of the
dynamical time $t_{D}$, the growth time $t_{G}$ is somewhat longer, 
corresponding more nearly to the time scale $t_{J}$ associated with 
${\Delta}J$.

This suggests strongly that the behaviour of ${\Delta}r$ and ${\Delta}v$ 
observed for the Plummer potential is associated with linear phase mixing. 
Because of finite number statistics, the same initial condition 
$({\bf r}_{0},{\bf v}_{0})$ in
different frozen-$N$ realisations of a Plummer potential will correspond to
somewhat different energies, the values of which are conserved in the
subsequent evolution. However, even neglecting discreteness effects, 
initially proximate orbits in a generic integrable potential will, if their
energies be unequal, tend to diverge linearly. For example, two orbits evolved
in a smooth Plummer potential with the same initial ${\bf r}$ but slightly
different values of ${\bf v}$ and, hence, slightly different energies, will
oscillate with somewhat different frequencies and, as a result, exhibit an
overall linear divergence. If, however, the orbits are evolved instead in the
potential associated with a constant density distribution, this is no longer
true. A constant density sphere corresponds to a harmonic potential, where
all orbits have the same unperturbed frequencies; and, for this reason, orbits
in the smooth potential with slightly different energies will not exhibit 
such a systematic divergence.

The fact that frozen-$N$ orbits look ``more nearly regular'' for large $N$
suggests that the chaos associated with discreteness effects
in the $N$-body problem should be viewed very differently from the chaos 
associated with a bulk nonintegrable potential. When evolved into the future,
two nearby chaotic initial conditions in such a potential tend to diverge
exponentially until they are separated by a distance comparable to the size
of the easily accessible ({\it i.e.,} not significantly impeded by the Arnold
web) connected phase space region to which the orbits are confined, a region
which tends, typically, to be macroscopic. By contrast, the scale associated
with chaos induced by discreteness effects in the $N$-body problem is 
distinctly microscopic, at least for comparatively large $N$. It would appear
that any single orbit with fixed energy can access a phase space region which
is in fact very large; but the chaos which it experiences is a superposition 
of short range effects with characteristic scale ${\ll}{\;}R_{sys}$.

\acknowledgments
The authors acknowledge useful discussions with Alexei Fridman, Salman Habib,
and Ilya Pogorelov. This research was supported in part by NSF AST-0070809 and 
by the Institute for Geophysics and Planetary Physics at Los Alamos National
Laboratory.

\vfill\eject
\pagestyle{empty}
\begin{figure}[t]
\centering
\centerline{
        \epsfxsize=8cm
        \epsffile{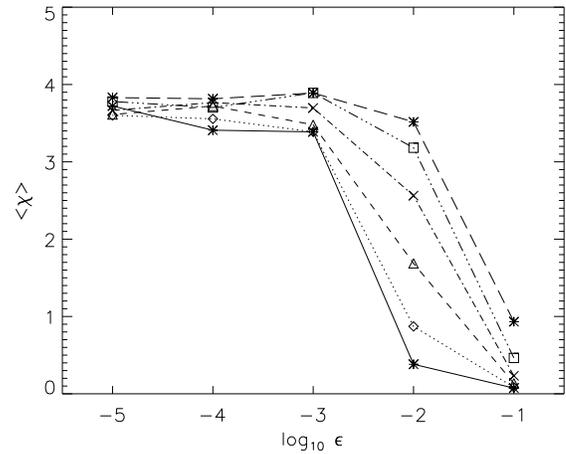}
           }
        \begin{minipage}{10cm}
        \end{minipage}
        \vskip -0.0in\hskip -0.0in
\caption{Mean short time Lyapunov exponent ${\langle}{\chi}{\rangle}$ as
a function of softening parameter ${\epsilon}$ for $N=10^{5}$ (solid line),
$N=10^{4.5}$ (dotted),
$N=10^{4}$ (dashed), $N=10^{3.5}$ (dot-dashed), $N=10^{3.0}$ (triple-dot
dashed), and $N=10^{2.5}$ (broad dashed). The integrations were all performed
for a single `typical' initial condition.}
\vspace{-0.0cm}
\end{figure}
\pagestyle{empty}
\begin{figure}[t]
\centering
\centerline{
        \epsfxsize=8cm
        \epsffile{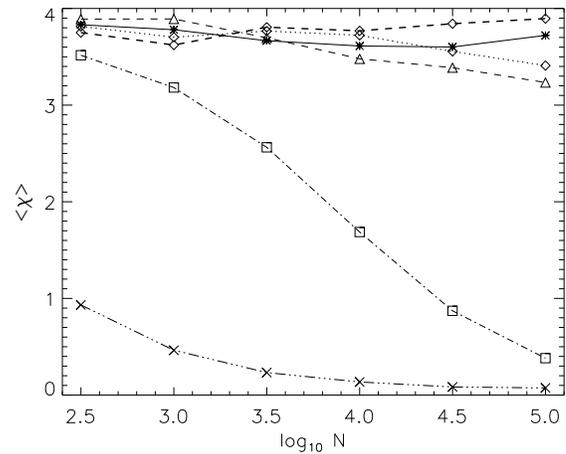}
           }
        \begin{minipage}{10cm}
        \end{minipage}
        \vskip -0.0in\hskip -0.0in
\caption{Mean short time Lyapunov exponent ${\langle}{\chi}{\rangle}$ as
a function of particle number $N$ for ${\epsilon}=10^{-5}$ (solid line),
${\epsilon}=10^{-4}$ (dotted),
${\epsilon}=10^{-3}$ (thin-dashed), ${\epsilon}=10^{-2}$ (dot-dashed), and
${\epsilon}=10^{-1}$ (triple-dot dashed), all computed for the initial 
condition used to generate FIG. 1. The short time ${\langle}{\chi}{\rangle}$ 
for a different initial condition corresponding to a smooth radial orbit, 
again evolved with ${\epsilon}=10^{-5}$, is 
indicated by the curve with thick dashes.}
\vspace{-0.0cm}
\end{figure}

\pagestyle{empty}
\begin{figure}[t]
\centering
\centerline{
        \epsfxsize=8cm
        \epsffile{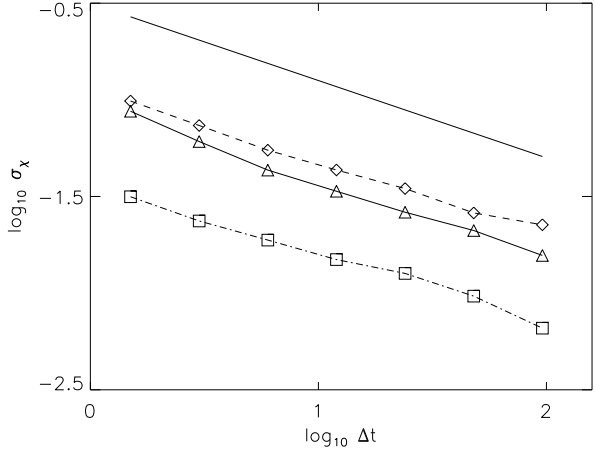}
           }
        \begin{minipage}{10cm}
        \end{minipage}
        \vskip -0.0in\hskip -0.0in
\caption{$\log_{10} {\sigma}_{\chi}({\Delta}t)$ as a function of
$\log_{10} {\Delta}t$ for three sets of simulations: $N=31623$ and ${\epsilon}
=0.0001$ (solid curve), $N=316$ and ${\epsilon}=0.0001$ (dashed curve), and
$N=316$ and ${\epsilon}=0.1$ (dot-dashed curve), all computed for the initial
condition used to generate FIG. 1. The thick solid line has a 
slope corresponding to $p=0.4$.}
\vspace{-0.0cm}
\end{figure}

\pagestyle{empty}
\begin{figure}[t]
\centering
\centerline{
        \epsfxsize=8cm
        \epsffile{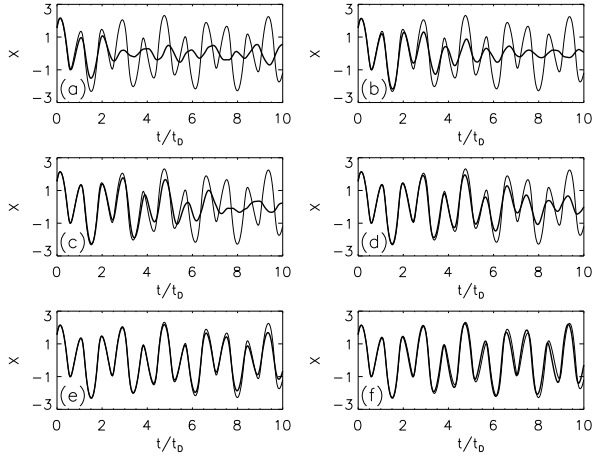}
           }
        \begin{minipage}{10cm}
        \end{minipage}
        \vskip -0.0in\hskip -0.0in
\caption{(a) The trajectory $x_{sm}(t)$ in the smooth potential (thin curve)
and the mean trajectory ${\langle}x(t){\rangle}$ (thick curve) derived from 
$20$ frozen-$N$ simulations with $N=316$ and ${\epsilon}=10^{-4}$, performed
for the initial condition used to generate FIG 1.
(b) The same for $N=1000$. (c) $N=3162$. (d) $N=10000$. (e) $N=31623$.
(f) $N=100000$.}
\vspace{-0.0cm}
\end{figure}

\pagestyle{empty}
\begin{figure}[t]
\centering
\centerline{
        \epsfxsize=8cm
        \epsffile{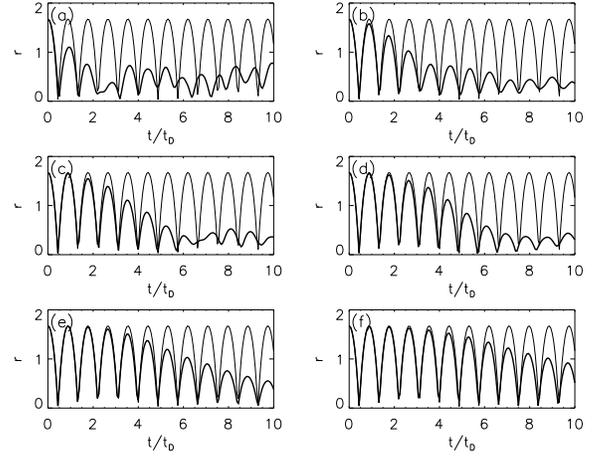}
           }
        \begin{minipage}{10cm}
        \end{minipage}
        \vskip -0.0in\hskip -0.0in
\caption{(a) The radial coordinate $r_{s}(t)$ in the smooth potential (thin 
curve) and the mean trajectory ${\langle}r(t){\rangle}$ (thick curve) derived 
from $20$ frozen-$N$ simulations with $N=316$ and ${\epsilon}=10^{-4}$, 
performed for an initial condition corresponding in the smooth potential to
a purely radial orbit.
(b) The same for $N=1000$. (c) $N=3162$. (d) $N=10000$. (e) $N=31623$.
(f) $N=100000$.}
\vspace{-0.0cm}
\end{figure}

\pagestyle{empty}
\begin{figure}[t]
\centering
\centerline{
        \epsfxsize=8cm
        \epsffile{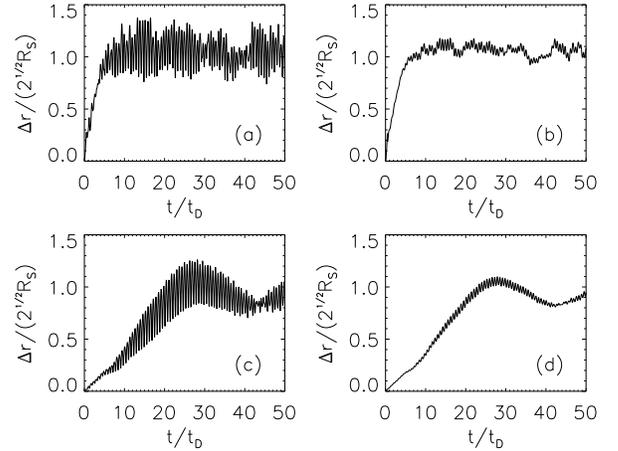}
           }
        \begin{minipage}{10cm}
        \end{minipage}
        \vskip -0.0in\hskip -0.0in
\caption{(a) The quantity ${\Delta}r/\sqrt{2R_{s}^{2}}$
for frozen-$N$ simulations with $N=1000$ and ${\epsilon}=10^{-4}$.
(b) The same data subjected to boxcar averaging over an interval $t=t_{D}$.
(c) and (d) The same for $N=100000$.}
\vspace{-0.0cm}
\end{figure}

\pagestyle{empty}
\begin{figure}[t]
\centering
\centerline{
        \epsfxsize=8cm
        \epsffile{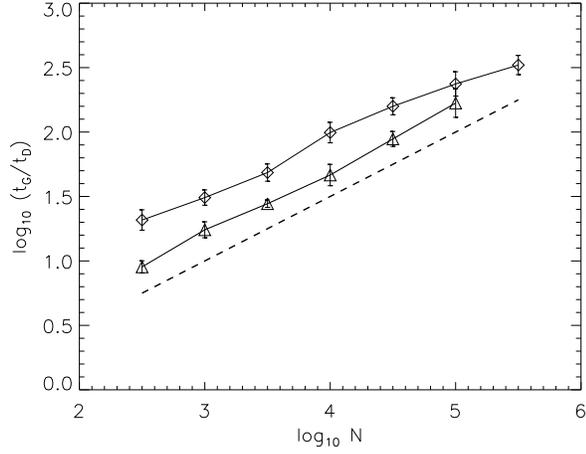}
           }
        \begin{minipage}{10cm}
        \end{minipage}
        \vskip -0.0in\hskip -0.0in
\caption{Best fit values of the time scale $t_{G}(N)$ associated with the
divergence of smooth and frozen-$N$ orbits for two different initial
conditions: ${\Delta}r/R_{s}=
{\Delta}v/V_{s}{\;}{\equiv}{\;}t/t_{G}.$ The dashed line has 
slope $1/2$, corresponding to an $N^{1/2}$ dependence.}
\vspace{-0.0cm}
\end{figure}

\pagestyle{empty}
\begin{figure}[t]
\centering
\centerline{
        \epsfxsize=8cm
        \epsffile{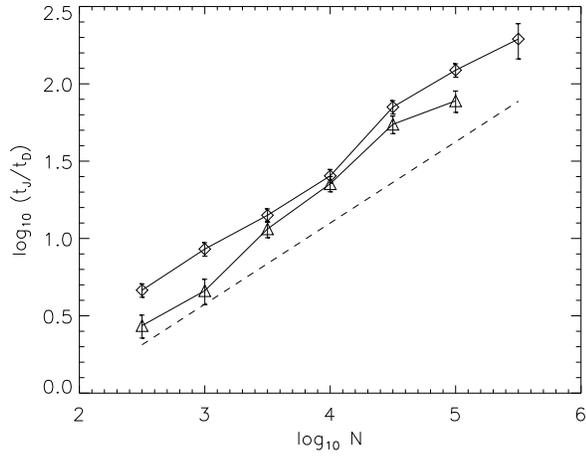}
           }
        \begin{minipage}{10cm}
        \end{minipage}
        \vskip -0.0in\hskip -0.0in
\caption{Best fit values of the time scale $t_{J}(N)$ associated with 
changes in angular momentum for frozen-$N$ orbits for two different initial
conditions: ${\Delta}J^{2}/J_{s}^{2}
{\;}{\equiv}{\;}t/t_{J}.$ The dashed line has slope $1/2$, corresponding to 
an $N^{1/2}$ dependence.}
\vspace{-0.0cm}
\end{figure}

\pagestyle{empty}
\begin{figure}[t]
\centering
\centerline{
        \epsfxsize=8cm
        \epsffile{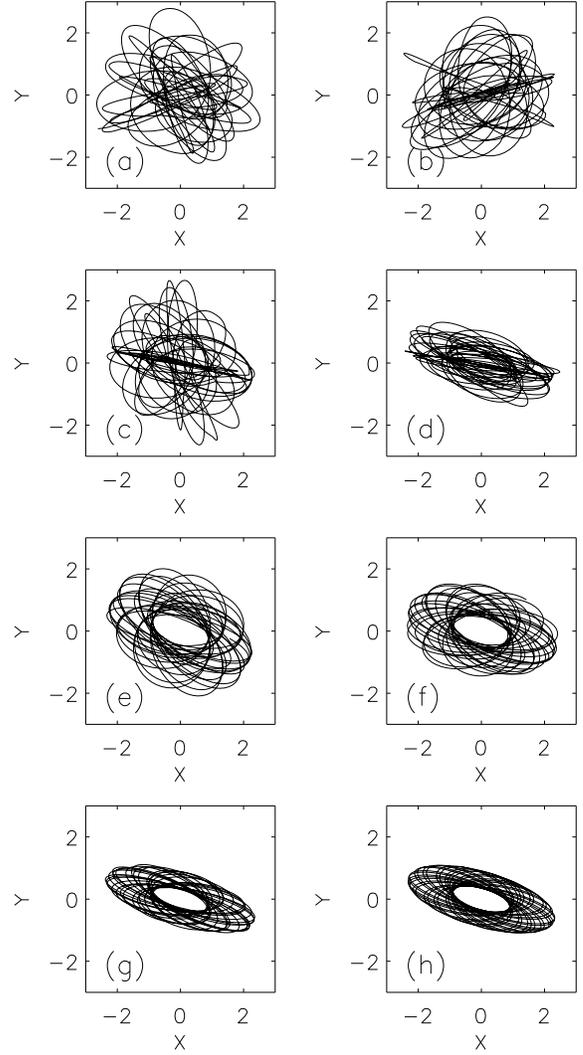}
           }
        \begin{minipage}{10cm}
        \end{minipage}
        \vskip -0.0in\hskip -0.0in
\caption{The $x$-$y$ projection of representative frozen-$N$ orbits generated 
from the same initial
condition, evolved for $t=25t_{D}$ with ${\epsilon}=10^{-5}$. (a) $N=316$.
(b) $N=1000$. (c) $N=3163$. (d) $N=10000$. (e) $N=31623$. (f) $N=100000$.
(g) $N=316228$. (h) The $x$-$y$ projection of the same initial condition 
evolved in the smooth potential}
\vspace{-0.0cm}
\end{figure}

\pagestyle{empty}
\begin{figure}[t]
\centering
\centerline{
        \epsfxsize=8cm
        \epsffile{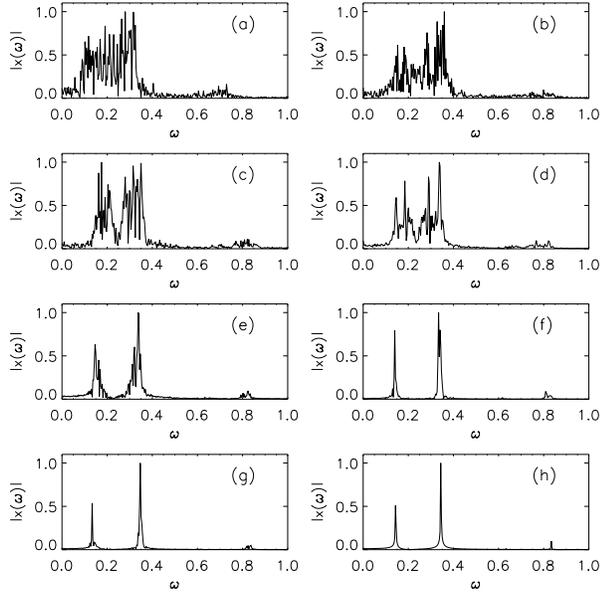}
           }
        \begin{minipage}{10cm}
        \end{minipage}
        \vskip -0.0in\hskip -0.0in
\caption{(a) The Fourier transformed $|x({\omega})|$ for one frozen-$N$
integration of the initial condition used to generate FIGURE 4, evolved with
${\epsilon}=10^{-5}$ and $N=316$. (b) The same for $N=1000$. (c) $N=3162$.
(d) $N=10000$. (e) $N=31623$. (f) $N=100000$. (g) $N=316228$. (h) 
$|x({\omega})|$ for a characteristic in the smooth potential with the same 
initial condition, with data recorded at the same intervals for the same
total integration time.}
\vspace{-0.0cm}
\end{figure}

\pagestyle{empty}
\begin{figure}[t]
\centering
\centerline{
        \epsfxsize=8cm
        \epsffile{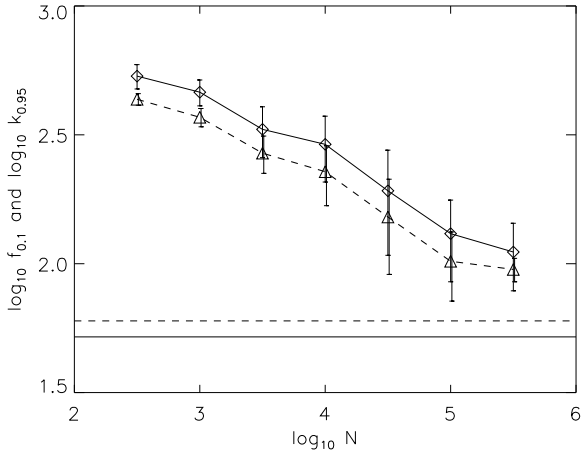}
           }
        \begin{minipage}{10cm}
        \end{minipage}
        \vskip -0.0in\hskip -0.0in
\caption{Two probes of the complexity of frozen-$N$ orbits for an ensemble of 
orbits with the same initial condition evolved with ${\epsilon}=10^{-5}$.
The solid curve exhibits $f_{0.1}$, the number of frequencies which have
power equal to at least 10\% of the power in the peak frequencies.
The dashed curve exhibits $k_{0.95}$, the
number of frequencies required to capture 95\% of the total power.
The vertical lines show $f_{0.1}$ and $k_{0.95}$ for a smooth characteristic
generated identically from the same initial condition, thus exhibiting
the intrinsic limitations associated with the discrete time series of data
points.}
\vspace{-0.0cm}
\end{figure}

\pagestyle{empty}
\begin{figure}[t]
\centering
\centerline{
        \epsfxsize=8cm
        \epsffile{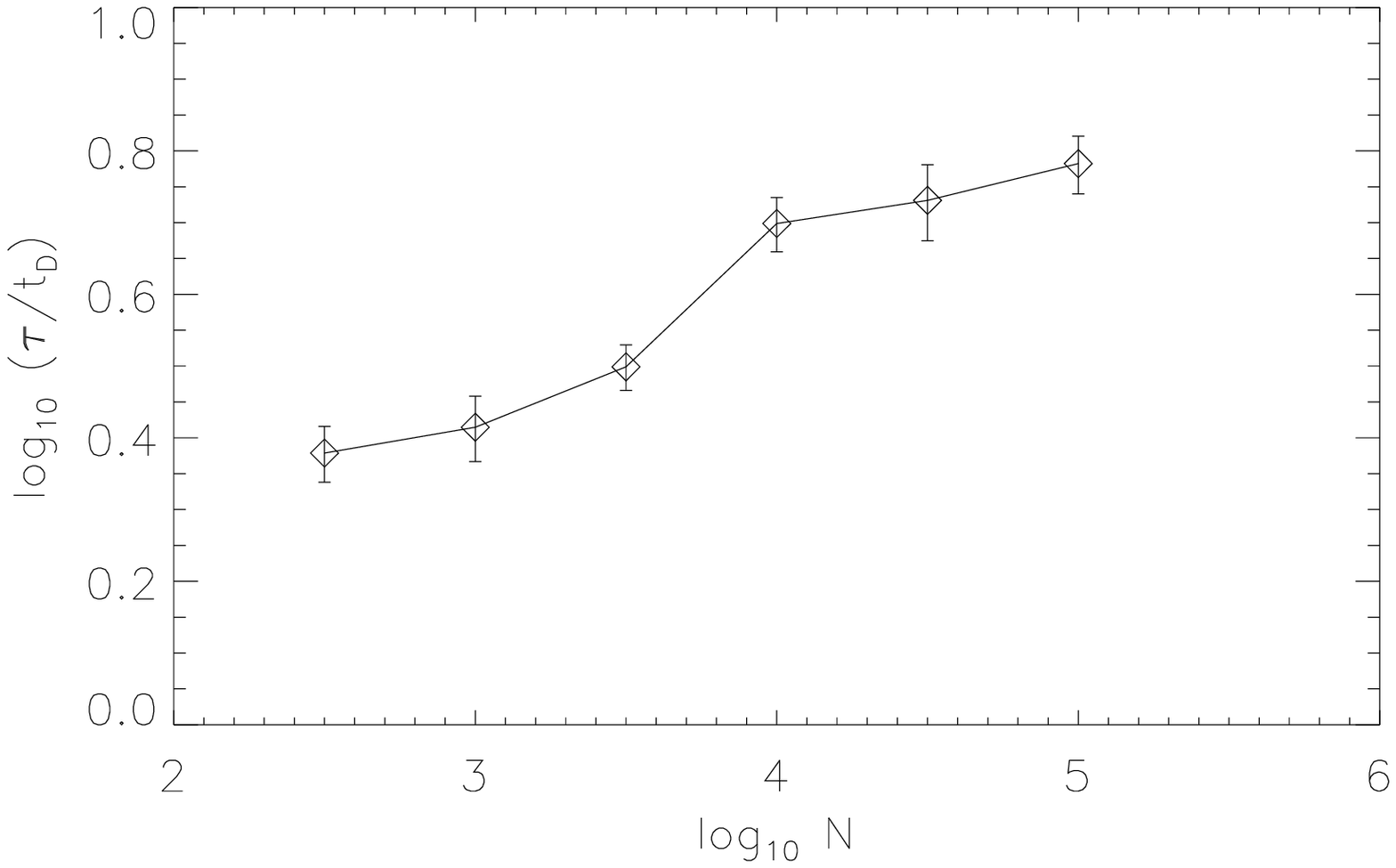}
           }
        \begin{minipage}{10cm}
        \end{minipage}
        \vskip -0.0in\hskip -0.0in
\caption{The mean time required for two frozen-$N$ orbits separated initially 
by a distance ${\delta}r=10^{-6}$ to achieve a macroscopic separation 
${\delta}r=1$.}
\vspace{-0.0cm}
\end{figure}
\vfill\eject

\vfill\eject


\begin{references}
\bibitem{1} R. H. Miller, Astrophys. J. {\bf 140}, 250 (1964).
\bibitem{2} H. E. Kandrup, Physica A {\bf 169}, 73 (1989) 
\bibitem{3} J. Goodman, D. Heggie, and P. Hut, Astrophys. J. {\bf 415}, 715
(1993)
\bibitem{4} Cf. H. E. Kandrup, M. E. Mahon, and H. Smith, Astrophys. J.
{\bf 428}, 458 (1994), and references cited therein.
\bibitem{5} Cf. J. Binney and S. Tremaine, {\it Galactic Dynamics} (Princeton
University Press, Princeton, 1987).
\bibitem{6} Cf. H. E. Kandrup and H. Smith, Astrophys. J. {\bf 374}, 255 
(1991).
\bibitem{7} S. J. Aarseth, M. H\'enon, and R. Wielen, Astron. Astrophys.
{\bf 37}, 183 (1974).
\bibitem{8}Cf. A. J. Lichtenberg and M. A. Lieberman, {\it Regular and Chaotic
Dynamics} (Springer, Berlin, 1992).
\bibitem{9}Cf. eq. (7.7) in H. E. Kandrup, Phys. Reports, {\bf 63}, 1 (1980).
\bibitem{10}C. Siopis and H. E. Kandrup, Mon. Not. Roy. Astro. Soc. {\bf 319},
43 (2000).
\bibitem{11}Cf. H. E. Kandrup, B. L. Eckstein, and B. O. Bradley, Astron.
Astrophys. {\bf 320}, 65 (1997), which discusses the pros and cons of
the admittedly somewhat simplistic probes of `complexity' used in this paper.
\bibitem{12}K. P. Rauch and S. Tremaine, New. Astron. {\bf 1}, 149 (1996).
\bibitem{13}A numerical investigation of the $N$-body problem in the presence
of a much larger central point mass (H. Smith, H. E. Kandrup, M. E. Mahon, and
C. Siopis, in {\it Ergodic Concepts in Stellar Dynamics}, edited by V. G.
Gurzadyan and D. Pfenniger, Springer Lectures Notes in Physics No. 430 
(Springer, New York, 1994), p. 158) suggests that, even if the central mass
$M_{BH}$ is much larger than the total mass $M=Nm$ of the individual particles,
the $N$-particle orbits continue to exhibit a sensitive dependence on initial
conditions. For example, for orbits in simulations with $M_{BH}=10M$, the
characteristic time scale $t_{*}$ associated with the exponential sensitivity
was found to be less than three times longer than the value of $t_{*}$ for 
$M_{BH}=0$.
\bibitem{14}M. Valluri and D. Merritt, in {\it The Chaotic Universe}, edited 
by R. Ruffini and V. G. Gurzadyan (World Scientific, New York, 1999).
\bibitem{15}Cf. S. Chandrasekhar, Rev. Mod. Phys. {\bf 15}, 1 (1943).
\end{references}
\end{document}